\documentstyle[psfig,aps,prl,twocolumn]{revtex}

\begin{document}
\draft


\author{M. K\"ohl$^{1,2,3}$\cite{email}, M. J. Davis$^{4,5}$, C. W. Gardiner$^6$, T. W.~H\"ansch$^{1,2}$, and
T. Esslinger$^{1,2,3}$}
\address{$^1$Sektion Physik,
Ludwig-Maximilians-Universit\"at, Schellingstr.\ 4/III, D-80799 Munich, Germany\\
$^2$Max-Planck-Institut f\"ur Quantenoptik, D-85748 Garching,
Germany\\
$^3$Institute for Quantum Electronics, ETH Z\"urich, CH-8093
Z\"urich, Switzerland\\
$^4$Clarendon Laboratory, Department of Physics, University of Oxford, Parks Rd, Oxford OX1 3PU, United Kingdom\\
$^5$Department of Physics, University of Queensland, St Lucia, QLD
4072, Australia\\
$^6$School of Chemical and Physical Sciences, Victoria University,
Wellington, New Zealand}
\date{Submitted June 29 2001}
\title{Growth of Bose-Einstein Condensates from Thermal Vapor}

\maketitle

\begin{abstract}
We report on a quantitative study of the growth process of
$^{87}$Rb Bose-Einstein condensates. By continuous evaporative
cooling we directly control the thermal cloud from which the
condensate grows. We compare the experimental data with the
results of a theoretical model based on quantum kinetic theory. We
find quantitative agreement with theory for the situation of
strong cooling, whereas in the weak cooling regime a distinctly
different behaviour is found in the experiment.

\end{abstract}

\pacs{03.75.Fi, 05.30.Jp}


\noindent

The non-equilibrium path to the formation of a Bose-Einstein
condensate \cite{bec} is one of the most intriguing physical
processes in ultra cold quantum gases. When the critical
temperature of this phase transition is crossed, the atoms
accumulate in the ground state of the system, which leads to the
characteristic dense core in the atomic distribution and to the
long-range phase coherence. It is an experimental challenge to
monitor this process in time, aiming for a precise understanding
of the formation of Bose-Einstein condensates. Experimental
studies of the growth of Bose-Einstein condensates have so far
either focused on verifying the effect of  bosonic stimulation
\cite{Miesner98} or on the case of attractive interactions between
the atoms, which gives rise to special dynamics \cite{Gerton2000}.
However, the key role of the thermal cloud, from which the
condensate grows, has not yet been addressed in experiments.

In a pioneering theoretical study Kagan {\it et al.}
\cite{Kagan1992} proposed the appearance of a quasi-condensate
with strong phase fluctuations prior to the formation of
long-range phase coherence. Stoof \cite{Stoof1997} presented the
idea of nucleation followed by a growth according to a kinetic
equation. A simple quantitative theory has been developed by
Gardiner {\it et al.} \cite{Gardiner1997,Gardiner1998} on the
basis of quantum kinetic theory. The key equation describing the
time dependence of the number $N_0$ of atoms in the condensate
reads
\begin{equation}
\dot N_0=2\,W^+(N_0)\left\{ \left(1-e^{\left[\mu_C(N_0)-\mu\right]/k_B T}\right)N_0+1\right\},
\label{simple}
\end{equation}
where $\mu_C(N_0)$ is the chemical potential of the condensate,
$\mu$ the chemical potential of the thermal vapor, and $W^+(N_0)$
an analytically known rate factor \cite{Gardiner1998}. Starting
from $N_0=0$ the condensate starts to grow at a finite rate, which
turns quickly into exponential growth, before it reaches its
asymptotic equilibrium value when the condensate chemical
potential approaches that of the thermal cloud. This equation
describes only the growth of the population in the ground state of
the system and it neglects certain scattering processes among
atoms in higher lying states. Especially in the initial period the
redistribution of population between higher lying states becomes
important and modifies the growth \cite{Gardiner1998}. When the
scattering terms are included one finds reasonable agreement with
the experiment described in \cite{Miesner98}, but qualitative and
quantitative discrepancies remain \cite{Gardiner1998,Davis2000}.
Moreover, equation (\ref{simple}) neglects the dynamics of the
thermal component by assuming its chemical potential to be
stationary. The influence of the dynamics of the thermal cloud was
included in the recent models \cite{Davis2000,Bijlsma2000}. A
purely numerical approach by Monte Carlo simulations is presented
in \cite{Wu1997}.

We have identified and studied different stages of formation when
the Bose-Einstein condensate grows from a thermal gas. In the
experiment a cloud of thermal atoms is prepared above the critical
temperature and then cooled into the quantum degenerate regime by
continuous evaporation for a duration of up to 6\,s. This strategy
allows us to observe the formation process in slow motion. A
distinct growth characteristic is revealed by directly controlling
the properties of the thermal vapor. In comparison, previous
experimental and most theoretical work addressed the situation in
which a short radio-frequency pulse is applied to remove hot
atoms, then turned off leaving the system to evolve toward thermal
equilibrium without any further forced evaporation.

For a quantitative understanding it is crucial to prepare a
well-defined initial state of the thermal cloud. As described
previously \cite{Esslinger98}, we load about $10^9$ atoms in the
$|F=1,m_F=-1\rangle$ hyperfine ground state into a magnetic
Quadrupole-Ioffe configuration trap (QUIC-trap). The trapping
frequencies are $\omega_\perp=2\pi\times 110$\,Hz in the radial
and $\omega_z=2 \pi \times 14$\,Hz in the axial direction. In the
magnetic trap we perform forced evaporative cooling for about
25\,s by continuously lowering an RF frequency to a final value of
$\nu_{RF,0}$=2120\,kHz. This prepares a purely thermal cloud of
$N_i$=(4.2$\pm$0.2)$\times 10^6$ atoms at a temperature of
$T_i$=(640$\pm$30)\,nK. The minimum of the trap is determined by
atom laser output coupling \cite{Bloch1999} and corresponds to a
radio-frequency of 1955\,kHz. The stability of the magnetic trap
enables us to approach the formation of the condensate very
slowly. The measured stability \cite{Koehl2001} and the good
reproducibility of the experiment show that the formation of the
condensate is not triggered by fluctuations of the magnetic field.
When operating the trap at full current, we have measured a slow
upwards drift of the bottom of the trapping potential of 5\,kHz/s,
which is due to thermally induced motion of the trap coils.

After this preparation of a purely thermal cloud, the atoms are
cooled through the BEC phase transition. The radio-frequency field
jumps abruptly to a lower frequency $\nu_{RF,1}$ and remains at
this value for a period of up to 6 seconds. This limits the depth
of the magnetic trap to $\epsilon_{cut}$, which is expressed in
units of $T_i$ by the evaporation parameter
$\eta=\epsilon_{cut}/k_B T_i$. Atoms with an energy larger than
$\epsilon_{cut}$ are lost from the trap due to spin flips into a
non-trapped state. Equilibration of the cloud by elastic
collisions leads to a decrease of temperature and the
Bose-Einstein condensate forms during this process.

The time evolution of the condensation is probed by releasing the
atoms from the magnetic trap after different evolution times, for
otherwise identical repetitions of the experiment. The atom cloud
is imaged by absorption imaging after 17\,ms of ballistic
expansion. Atom numbers are extracted from the absorption images
by two-dimensional fits to the atomic density distribution. The
thermal cloud is fitted by a Bose-Einstein distribution under the
assumption of zero chemical potential, and the condensed fraction
is fitted by a Thomas-Fermi distribution. The error in the
determination of the atom number is estimated to be below 10\%.
Here, the ''condensate number'' refers to the integral under the
Thomas-Fermi profile. Temperatures are determined from a Gaussian
fit to the far out wings of the thermal atom cloud. The
statistical error on the temperature determination is 5\%, whereas
the systematic error is 10-15\%.

The initially prepared atomic sample corresponds to $\eta=6$,
where the gravitational sag of the atom cloud must be taken into
account. Gravity pulls the atomic cloud away from the center of
the magnetic trap and thereby reduces the dimensionality and the
efficiency of evaporative cooling. We have performed experiments
in the range $0.75<\eta<4.8$, corresponding to RF frequencies
between 1990\,kHz and 2095\,kHz. For $\eta>4.6$ we could not
obtain Bose-Einstein-condensates within the 6\,s period of the
measurement due to the low efficiency of the cooling. For
$\eta=0.75$ we did not observe the formation of a condensate since
the atom number of the initial thermal cloud was reduced too much
for evaporative cooling to proceed.

\begin{figure}
\centerline{\psfig{file=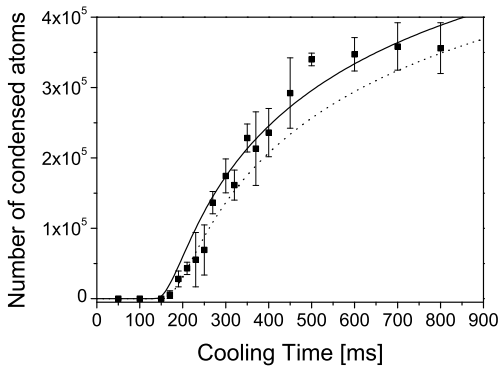,angle=0,width=0.85\columnwidth}}
\caption{Growth curve of a Bose-Einstein condensate for an
evaporation parameter $\eta=1.4$. The lines are the results of the
numerical simulation of the growth for the starting conditions
$N_i=4.4\times 10^6$ and $T_i=610$\,nK (solid) and $N_i=4.2\times
10^6$ and $T_i=610$\,nK (dotted). Every data point is averaged
over 3 identical repetitions of the experiment with statistical
errors shown by the bars.} \label{fig1}
\end{figure}

Figure \ref{fig1} shows the measured data for an evaporation
parameter $\eta=1.4$. Every data point is an average over three
measurements and the error bars show the statistical error. The
final temperature of the cloud is (220$\pm$20)\,nK and the final
condensate fraction is ($55\pm7$)\%. The solid line is a numerical
calculation of the condensate growth according to a model based on
quantum kinetic theory, where the influence of the thermal atoms
is taken into account \cite{Davis2000}. The model was improved to
include the continuous evaporation of atoms, the gravitational sag
of the atomic cloud, the experimentally observed magnetic field
drift, and three-body loss of atoms from the condensate
\cite{Davis2001a}.  We find quantitative agreement between the
model and the measured data without any free parameters entering
into the calculation, which suggests that the model describes this
regime of the formation of Bose-Einstein condensates sufficiently.
The best agreement between data and theory has been obtained for
the calculation with the starting conditions $N_i=4.4\times 10^6$
and $T_i=610$\,nK or $N_i=4.2\times 10^6$ and $T_i=610$\,nK, both
of which lie within the range of the experimental uncertainties.

\begin{figure}
\centerline{\psfig{file=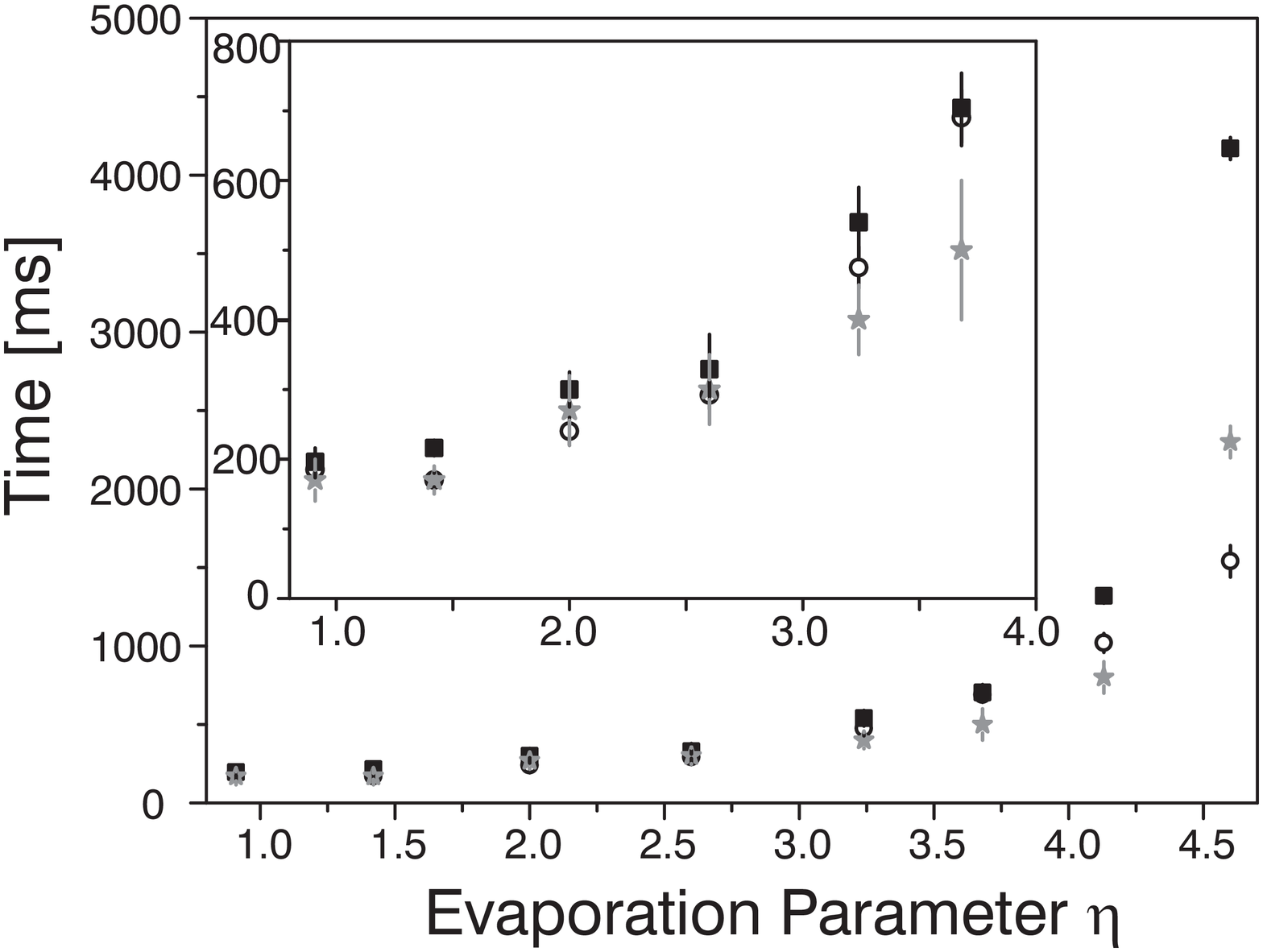,angle=0,width=0.85\columnwidth}}
\caption{Initiation times for the growth of Bose-Einstein
condensates. The squares show the measured initiation times
according to equation (\ref{exp}), the open circles show the
times, when a condensate is detected for the first time, and the
gray stars show the time the condensate population exceeds $10^4$
as obtained from theoretical simulations. The inset shows an
enlargement of the results for small $\eta$.} \label{fig2}
\end{figure}

\begin{figure}
\centerline{\psfig{file=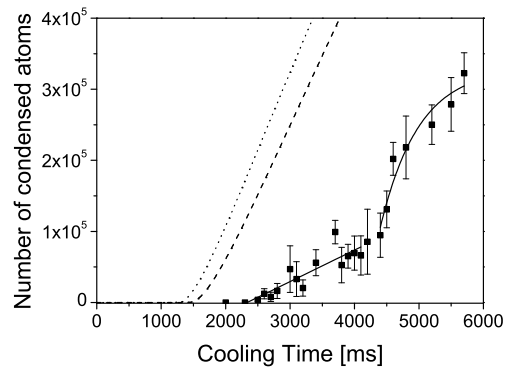,angle=0,width=0.85\columnwidth}}
\caption{Condensate growth curve for $\eta=4.6$.  The kink in the
growth curve is clearly seen. It occurs at a condensate fraction
of $N_0/N=3.5\%$. The linear growth rate is inferred from a linear
fit (straight line). The relaxational part of the growth curve is
fitted by equation (\ref{exp}) (curved line). The initial rate in
this regime is eight times larger than the linear growth rate.
Every data point is averaged over 4 identical repetitions of the
experiment with statistical errors shown by the bars. The results
of the numerical calculation are shown for the same initial
parameters as in fig. \ref{fig1}: $N_i=4.4\times 10^6$ and
$T_i=610$\,nK (dashed) and $N_i=4.2\times 10^6$ and $T_i=610$\,nK
(dotted). } \label{fig4}
\end{figure}

We have compared the time of the onset of Bose-Einstein
condensation after the beginning of the final radio frequency
cooling with the calculated curves and find excellent agreement
for the situation of strong cooling (see Fig. \ref{fig2}). To
quantify the initiation time $t_i$ for the measured data we have
fitted the function
\begin{equation}
N_0(t)=N_{0,f}\cdot\left(1-e^{-\Gamma(t-t_i)}\right), \label{exp}
\end{equation}
which describes the final stage of the condensate growth towards
equilibrium (see Fig. \ref{fig4}). $\Gamma$ describes a relaxation
rate towards equilibrium. This functional form was suggested in
Ref. \cite{Holland1997} and subsequently found to be a reasonable
approximation to more sophisticated calculations
\cite{Gardiner1998,Bijlsma2000}. To account for the range of
validity of the equation we have restricted the fitting range to
condensate numbers larger than 30\% of $N_{0,f}$, where we have
obtained good agreement with the data. For low values of the
evaporation parameter, i.e. rapid cooling, the initiation time
approaches a minimum of 200\,ms. For the example shown in figure
\ref{fig1} we obtain an initiation time of $t_i=(215\pm10)$\,ms.
This time corresponds to $30\,\tau_0$, where $\tau_0=(n\sigma
v)^{-1}$ is the classical collision time in a thermal gas ($n$:
mean atomic density, $\sigma$: scattering cross section, $v$:
average thermal velocity). For the homogeneous Bose gas the first
stage of formation was predicted to be on the order of $\tau_0$
\cite{Kagan1992}. Early numerical calculations have confirmed that
this should also be valid for the case of a harmonic confining
potential, provided that all scattering processes are included
\cite{Gardiner1998,Bijlsma2000}. Most theoretical treatments
require sufficient ergodicity, meaning that ergodic mixing occurs
on the order of a few collision times. At temperatures below a few
$\mu$K this does not describe the experimental situation.
Evaporation in the magnetic trap becomes one-dimensional due to
the gravitational sag of the atom cloud. This leads to a
significantly lower evaporation rate \cite{Surkov1996} and the
ergodic mixing times can be on the order of several ten collision
times \cite{Wu1997}.

The measured initiation times show a pronounced divergence at
$\eta\simeq4.6$ and become much longer than the ergodic mixing
time of some ten $\tau_0$. Very slow cooling through the phase
transition does not only affect the initiation time but it also
changes the characteristics of the growth curve. We detected a
slow, approximately linear growth of a condensate prior to the
initiation time, determined from equation (\ref{exp}). The open
circles in figure \ref{fig2} indicate when a condensate is
observed for the first time and the gray stars show the time the
condensate population exceeds $10^4$ as obtained from theoretical
simulations of the growth experiments \cite{Davis2001a}.

In figure \ref{fig4} the two distinct stages of condensate
formation can be seen for the case of slow cooling ($\eta=4.6$):
first, there is a approximately linear growth at constant rate of
$\dot N_0=(44\pm8)\times10^3$\,s$^{-1}$, followed by a very rapid
growth with an initial rate 8 times larger. The slow linear growth
of the condensate indicates a less pronounced bosonic stimulation
during this stage of the condensate formation. The numerical
calculation does not reproduce the measured condensate growth data
accurately in this regime. The calculation rather shows a growth
rate of the same order of magnitude as the second stage of the
observed growth, suggesting that the model works well if the
system is sufficiently far below the transition temperature. For
the thermal cloud the calculated temperatures agree with the
experimentally determined values. We deduce that a possible
heating rate of the atoms in the trap is below our sensitivity of
5~nK/s and thus probably negligible. The measured decrease in the
total atom number for long time scales seems to be slightly larger
than expected from the simulations. Whereas three body losses in
the condensate are included in the calculations, we neglect three
body recombination involving thermal atoms. The factor of 6
enhancement for these processes is negated by the much lower
density of the thermal cloud cloud: for the given trap parameters
and a temperature of 300\,nK one obtains a lifetime of more than
40\,s \cite{Burt1997,remark}. Similarly, the background limited
lifetime of the magnetic trap is at least 40\,s.

We attribute the discrepancy between the theoretical model and the
measured condensate numbers for slow cooling to approximations in
the theoretical model which for example do not take into account
coherence effects in the scattering process between the atoms.
Possibly, the two-stage growth can be explained by the formation
of quasi-condensates which have been predicted as an early stage
in the formation process of a Bose-Einstein condensate
\cite{Kagan1992}. Quasi-condensates exhibit strong phase
fluctuations that die out as the system grows. These phase
fluctuations are due to a population of low-lying excited states
in the trap. Density fluctuations are suppressed due to their
comparatively large excitation energy and therefore the density
exhibits a Thomas-Fermi distribution. The growth rate could then
be significantly smaller than expected from the assumption of
Bose-stimulation. Indeed, we find the growth rate to be almost one
order of magnitude smaller than calculated from our model based on
quantum kinetic theory. The time scale for both stages of the
evolution is longer than the ergodic mixing time. The system is
closer to equilibrium compared to the case of rapid cooling, i.e.
small values of the evaporation parameter $\eta$. However, the
kink in the growth curves clearly shows that the condensate is not
formed adiabatically. In this case the atom number should increase
according to $N_0=N(1-(T/T_c)^3)$ when the temperature is lowered
smoothly.

It was recently found that 3D Bose-Einstein condensates can
exhibit strong phase fluctuations in elongated traps
\cite{Petrov2001a}. The phase fluctuations are characterized by
the parameter $\delta_L^2=(T/T_c)(N/N_0)^{3/5} \delta_c^2$, where
$\delta_c^2$ is a factor that mainly depends on the trap geometry
and atom number. Its value is about 0.15 for our experiment. The
kink in the growth curves is observed at temperatures higher than
0.9\,$T_c$, thus large phase fluctuations $\delta_L^2>1$ can be
expected in the regime $N_0/N<4\%$. We find the transition from
linear growth to relaxational growth for slow cooling always in
the range $1.5\%<N_0/N<3.5\%$, which is compatible with the
theoretical expectation. When the phase fluctuations have died
out, i.e. $\delta_L^2\ll 1$, a true Bose-Einstein condensate
forms. This stage of the evolution then shows the same growth
characteristics as observed for rapid cooling. So far calculation
of the kinetics of the growth process have not taken into account
the large aspect ratio of cigar shaped traps. Therefore axial
phase fluctuations may be not included in these models. Evidence
for axial phase fluctuations in a steady state of a very elongated
condensate is reported in \cite{Dettmer2001a}.


In conclusion, we have investigated the influence of the thermal
atom cloud on the growth of a Bose-Einstein condensate. The
measured data has been compared to a numerical model based on
quantum kinetic theory. We find the minimum initiation time for
condensate growth to be 30 times the classical collision time, in
quantitative agreement with our calculations. This comparatively
long timescale is presumably due to slow ergodic mixing of the
trapped atomic cloud. Very slow cooling of the cloud enables us to
observe a two stage process during the formation of a
Bose-Einstein condensate, which is possibly beyond the
approximations made in the model.

We would like to thank I. Bloch for help with the data analysis
software, W. Eckardt for assistance with the laser system, Yu.
Kagan, W. Ketterle, and W. Zwerger for helpful comments and DFG
for financial support of the experimental work. MJD was supported
by the UK EPSRC, CWG was supported by the Royal Society of New
Zealand under the Marsden Fund Contract No. PVT-902.

\begin{references}
\bibitem[*]{email} email: koehl@iqe.phys.ethz.ch
\bibitem{bec} M. H. Anderson {\it et. al.}, Science {\bf269}, 198 (1995); K. B. Davis {\it et. al.}, Phys. Rev. Lett.
{\bf75}, 3969 (1995); C. C. Bradley {\it et. al.}, Phys. Rev.
Lett. {\bf75}, 1687 (1995) and {\bf78}, 985 (1997).
\bibitem{Miesner98} H. J. Miesner {\it et al.}, Science, {\bf 279},
1005 (1998).
\bibitem{Gerton2000} J. M. Gerton, D. Strekalov, I. Prodan, and R.
G. Hulet, Nature {\bf 408}, 692 (2000).
\bibitem{Kagan1992} Yu. M. Kagan, B. V. Svistunov, and G. V.
Shlyapnikov, Sov. Phys. JETP {\bf 75}, 387 (1992).
\bibitem{Stoof1997} H. T. C. Stoof, Phys. Rev. Lett. {\bf 66}, 3148 (1991); H. T. C. Stoof, Phys. Rev. Lett. {\bf 78},
768 (1997).
\bibitem{Gardiner1997} C. W. Gardiner, P. Zoller, R. J. Ballagh, and M. J.
Davis, Phys. Rev. Lett. {\bf 79}, 1793 (1997).
\bibitem{Gardiner1998} C. W. Gardiner, M. D. Lee, R. J. Ballagh,
M. J. Davis, and P. Zoller, Phys. Rev. Lett. {\bf 81}, 5266 (1998); M. D. Lee and C. W. Gardiner,
Phys. Rev. A {\bf 62}, 033606 (2000).
\bibitem{Davis2000} M. J. Davis, C. W. Gardiner, and R. J.
Ballagh, Phys. Rev. A {\bf 62}, 063608 (2000).
\bibitem{Bijlsma2000} M. J. Bijlsma, E. Zaremba, and H. T.
C. Stoof, Phys. Rev. A {\bf 62}, 063609 (2000).
\bibitem{Wu1997} H. Wu, E. Arimondo, and C. J. Foot, Phys. Rev. A
{\bf 56}, 560 (1997).
\bibitem{Esslinger98} T. Esslinger, I. Bloch, and T. W.
H\"ansch, Phys. Rev. A {\bf58}, R2664 (1998).
\bibitem{Bloch1999} I. Bloch, T. W. H\"ansch, and T. Esslinger, Phys. Rev. Lett.
{\bf82}, 3008 (1999).
\bibitem{Koehl2001} M. K\"ohl, T. W. H\"ansch, and T.
Esslinger, Phys. Rev. Lett. {\bf 87}, 160404 (2001).
\bibitem{Davis2001a} M. J. Davis and C. W. Gardiner, cond-mat/0111444.
\bibitem{Surkov1996} E. L. Surkov, J. T. M. Walraven and G. V.
Shlyapnikov, Phys. Rev. A {\bf 53}, 3403 (1996).
\bibitem{Holland1997} M. Holland, J. Williams, and J. Cooper,
Phys. Rev. A {\bf 55}, 3670 (1997).
\bibitem{Burt1997} E. A. Burt {\it et al.}, Phys. Rev. Lett. {\bf
79}, 337 (1997); J. S\"oding {et al.}, Appl. Phys. B {\bf 69}, 257
(1999).
\bibitem{remark} This calculation does not take into account the
repulsion of the thermal cloud by the growing condensate and thus
rather overestimates the three body recombination rate.
\bibitem{Petrov2001a} D. S. Petrov, G. V. Shlyapnikov, and J. T.
M. Walraven, Phys. Rev. Lett. {\bf 87}, 050404 (2001).
\bibitem{Dettmer2001a} S. Dettmer {\it et al.}, Phys. Rev. Lett. {\bf 87}, 160406 (2001).


\end{references}
\end{document}